\documentclass[traditabstract]{aa} 
\usepackage{natbib} 
\usepackage{graphicx}
\usepackage{txfonts}
\usepackage{epsfig}
\def\xte{XTE\,J1743-363}
\def\inte{{\em INTEGRAL}}
\def\xmm{{\em XMM-Newton}}

\def\chan{{\em Chandra}}

\def\beppo{{\em BeppoSAX}}
\def\rxte{{\em RXTE}}
\def\swift{{\em Swift}}

\begin{document}

   \title{\xmm\ and \swift\ observations of \xte\ }

   \author{E. Bozzo 
          \inst{1}
        \and P. Romano
          \inst{2}
        \and  C. Ferrigno
           \inst{1}   
          \and S. Campana
         \inst{3} 
        \and M. Falanga 
           \inst{4} 
         \and G. Israel 
         \inst{5}
         \and  R. Walter
           \inst{1}  
          \and L. Stella
         \inst{5}         
          }
   \institute{ISDC Data Centre for Astrophysics, department of Astronomy, University of Geneva, Chemin d’Ecogia 16,
             CH-1290 Versoix, Switzerland; \email{enrico.bozzo@unige.ch}
         \and
         INAF, Istituto di Astrofisica Spaziale e Fisica Cosmica - Palermo, Via U. La Malfa 153, I-90146 Palermo, Italy
         \and
         INAF - Osservatorio Astronomico di Brera, Via Bianchi 46, I-23807 Merate (LC), Italy 
        \and
        International Space Science Institute (ISSI) Hallerstrasse 6, CH-3012 Bern, Switzerland 
        \and
        INAF - Osservatorio Astronomico di Roma, Via Frascati 33, I-00044 Roma, Italy
             }
   
   \date{Submitted: 2013 February 2; Accepted 2013 June 6}

  \abstract{\xte\ is a poorly known hard X-ray transient, that displays short and intense flares similar to those observed
  from Supergiant Fast X-ray Transients. The probable optical counterpart shows spectral properties similar to those of an M8 III giant, 
  thus suggesting that \xte\ belongs to the class of the Symbiotic X-ray Binaries. 
  In this paper we report on the first dedicated monitoring campaign of the source in the soft X-ray range with \xmm\ and \swift\,/XRT. 
  These observations confirmed the association of \xte\ with the previously suggested M8 III giant and the classification of the source as a member of the 
  Symbiotic X-ray binaries. In the soft X-ray domain, \xte\ displays a high absorption  
  ($\sim$6$\times$10$^{22}$~cm$^{-2}$) and variability on time scales of hundreds to few thousand seconds, typical of wind accreting systems. 
  A relatively faint flare (peak X-ray flux 3$\times$10$^{-11}$~erg/cm$^2$/s) lasting $\sim$4~ks is recorded during the \xmm\ observation and interpreted 
  in terms of the wind accretion scenario.}

  \keywords{gamma rays: observations -- X-rays: individuals: XTE\,J1743-363}

   \maketitle

\section{Introduction}
\label{sec:intro}

\xte\ was discovered by \citet{markwardt99} during \rxte\ monitoring
observations of the Galactic-center region. At that time, the source was detected undergoing episodes 
of enhanced X-ray activity reaching fluxes of 15~mCrab and showing variability with time scales as short as $\sim$1~min. 
The source was also observed undergoing intense short outbursts with \inte\ and was thus classified as 
a possible Supergiant Fast X-ray Transient \citep[SFXT, see e.g.,][]{sguera06}. 
The brightest outburst from the source observed with \inte\ reached about 40~mCrab in the 20-60~keV 
energy band and lasted $\sim$2.5~h. 
Before the present work, \xte\ was not studied with focusing X-ray telescopes in the soft X-ray domain (0.3-12~keV) and the  
position of the source could be securely measured only with an accuracy of a few arcmin.   
A possible sub-arcmin localization of the source was proposed by \citet{ratti10} using an archival \chan\ observation. In those data, 
about 11 photons were serendipitously recorded from a location comprised in the relatively large \inte\ error circle 
around the position of \xte.\ Based on this, \citet{smith12} proposed a probable  
optical counterpart which displays typical spectral characteristics of an M8III giant, 
suggesting that \xte\ could be a new member of the Symbiotic X-ray Binary (SyXB) class rather than a SFXT   
(though some uncertainties still prevent a firm conclusion about the precise spectral type of the companion).  
\citet{smith12} reported also a complete re-analysis of all \rxte\,/PCA data of the source and showed that 
its X-ray flux has continuously decreased during the past $\sim$13~yrs. Such a behavior was not observed before in a SFXT.  

We report in this paper on the first pointed observations of \xte\ with focusing X-ray instruments. 
These include a deep pointing with \xmm\ (total exposure time 42~ks) and a 3~month-long monitoring campaign performed 
with \swift\,/XRT (total on-source exposure time 79~ks).

\section{\xmm\ data analysis and results}
\label{sec:data}

\begin{figure}
\centering
\includegraphics[scale=0.13]{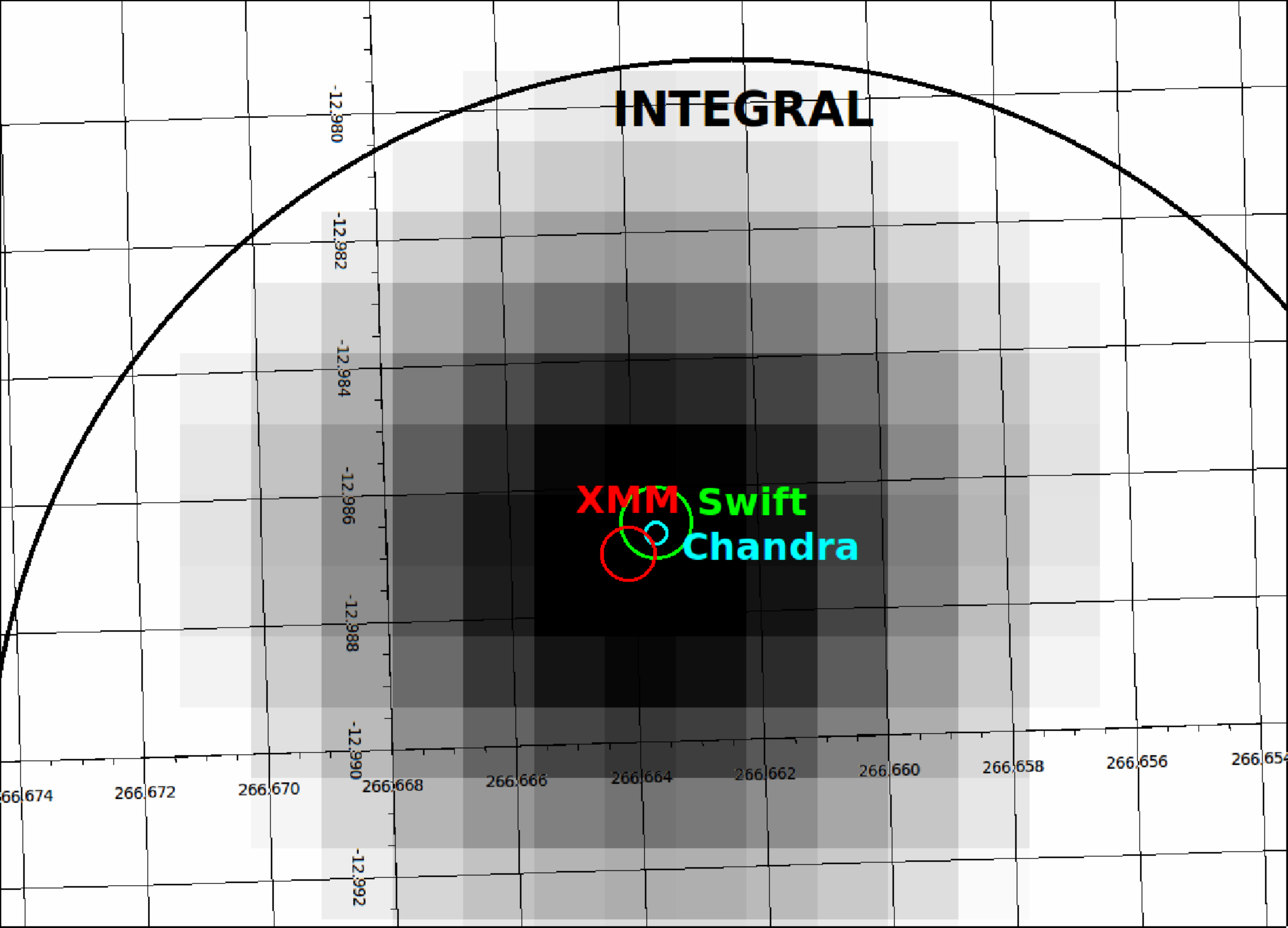}
\caption{The refined XMM position (1.5'' at 68\% c.l.) of the source reported in the present work is compatible with the \chan\ 
position of the faint source detected by \citet[0.6'' at 90\% c.l.;][]{ratti10} and suggested to be the quiescent counterpart to \xte.\ 
We also show the \inte\ position \citep[0.7' at 90\% c.l.;][]{bird10} and the \swift\,/XRT position (2.3'' at 90\% c.l.; 
see Sect.~\ref{sec:swift}).}     
\label{fig:position} 
\end{figure}
\xmm\ observed \xte\ for about 42~ks starting on 2012 February 29 18:31 UT. 
The EPIC-pn was operated in full-frame, the EPIC-MOS1 in small window and 
the MOS2 in timing mode. 
Observation data files (ODFs) were processed to produce calibrated
event lists using the standard \xmm\ Science Analysis
System (v. 12.0.1). We used the {\sc epproc} and {\sc
emproc} tasks to produce cleaned event files from the Epic-pn and MOS
cameras, respectively. Epic-pn and Epic-MOS event files were filtered in the 0.3-12~keV and 
0.5-10~keV energy range, respectively. High background time intervals were excluded using standard 
techniques\footnote{http://xmm.esac.esa.int/sas/current/documentation/threads/ EPIC\_filterbackground.shtml.}.   
The effective exposure time was of 33~ks for the EPIC-pn and 40~ks for the MOS cameras. 
Lightcurves and spectra of the source and background  
were extracted by using regions in the same CCD for the MOS and from adjacent CCDs  
at the same distance to the readout node for the EPIC-pn\footnote{See http://xmm.esac.esa.int/sas/current/documentation/threads.}.  
The difference in extraction areas between source and background was accounted for by using the SAS {\sc
backscale} task for the spectra and {\sc lccorrr} for the lightcurves. 
All EPIC spectra were rebinned before fitting in order to have at least 15-25 counts per bin 
(depending on the statistics of each spectrum) and, at the same 
time, prevent oversampling of the energy resolution by more than a factor of three. 
Where required, we barycenter-corrected the photon arrival times in the EPIC event files 
with the {\em barycen} tool. 
Throughout this paper, uncertainties are given at 90\%~c.l., unless stated otherwise.
\begin{figure}
\centering
\includegraphics[scale=0.35,angle=-90]{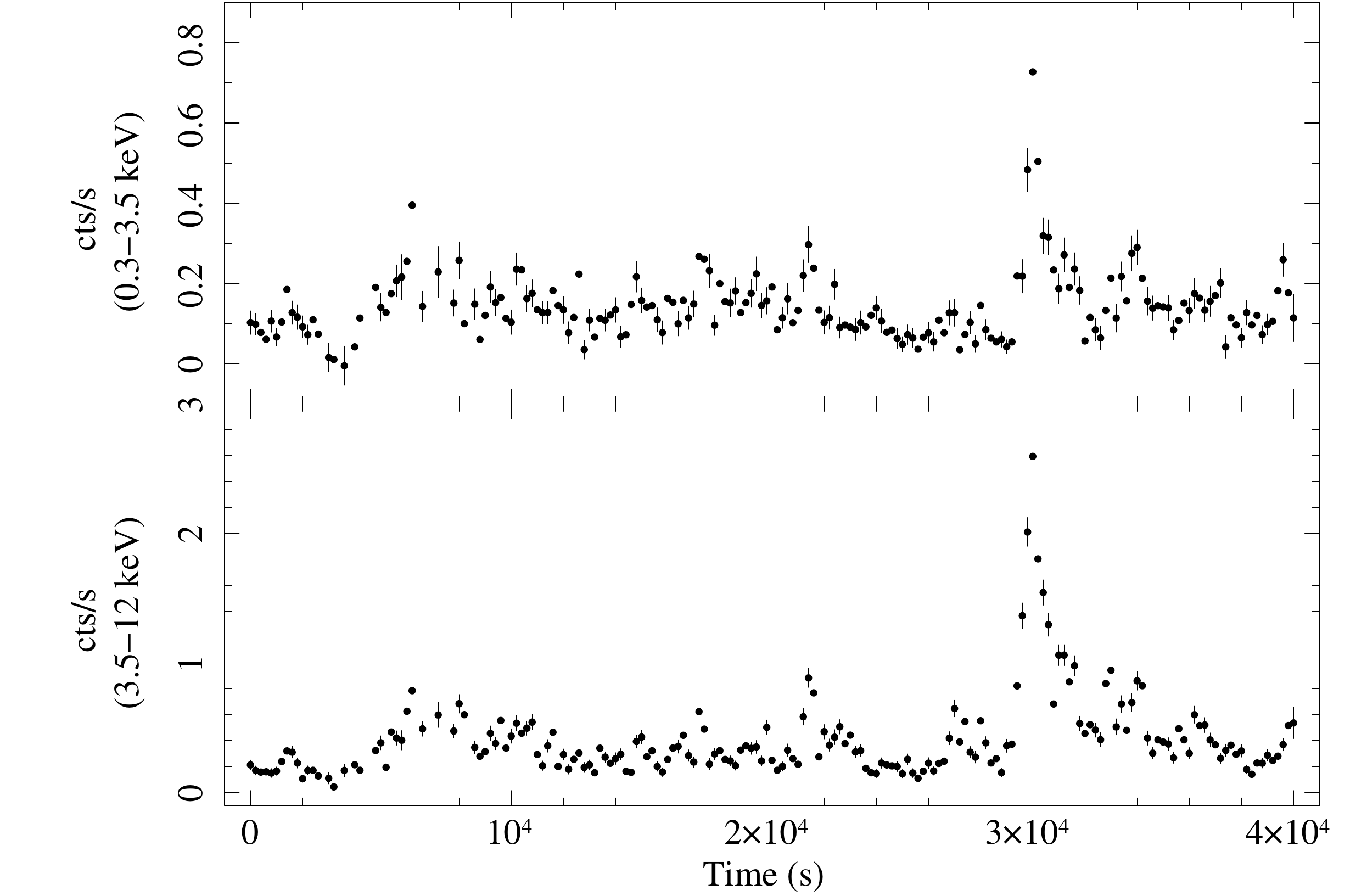}
\caption{EPIC-pn lightcurve of \xte\ in the 0.3-3.5~keV (upper panel) and 3.5-12~keV (lower-panel) 
energy band. The start time is 55986.8017~MJD and the bin time is 200~s.}    
\label{fig:lcurve} 
\end{figure}
\begin{figure}
\centering
\includegraphics[scale=0.36]{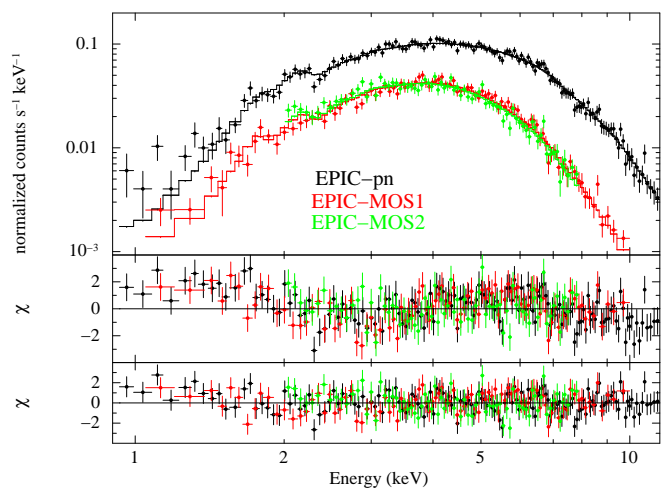}
\caption{EPIC-pn (black), EPIC-MOS1 (red), and EPIC-MOS2 (green) spectra of \xte\ extracted by using the entire exposure time available from the \xmm\ observation. 
The best fit is obtained here with an absorbed BMC model. The residuals from this fit are shown in the 
bottom panel. The middle panel shows the residuals from a fit with a simple power-law model.}     
\label{fig:totalspectrum} 
\end{figure}
The best position of \xte\ determined by the EPIC-pn and MOS1 is at $\alpha_{\rm J2000}$=17$^{\rm h}$43$^{\rm m}$01$^{\rm s}$.44 and 
$\delta_{\rm J2000}$=-36$^{\circ}$22$'$23$''$.16, with an associated uncertainty of 1.5~arcsec.  
at 68\% c.l.\footnote{See http://xmm.esac.esa.int/sas/current/documentation/threads/ src\_find\_thread.shtml and http://xmm2.esac.esa.int/docs/documents/ CAL-TN-0018.ps.gz.} 
In Fig.~\ref{fig:lcurve} we show the EPIC-pn background-subtracted source lightcurve in two energy bands. 
A flare from the source is observed about 30~ks after the beginning of the observation. 
The averaged EPIC-pn, MOS1 and MOS2 spectra of the observation are shown in Fig.~\ref{fig:totalspectrum}.   
A simultaneous fit to all spectra\footnote{The EPIC-MOS2 spectrum (operated in timing mode) showed evident instrumental residuals below 2.0~keV and above 8.0~keV, not present in the other 
spectra of the source. These data were thus discarded for further analysis.} with a simple absorbed power-law model gave an absorption column density $N_{\rm H}$=(6.1$\pm$0.2)$\times$10$^{22}$~cm$^{-2}$, 
a photon index $\Gamma$=1.51$\pm$0.05, and a poor $\chi^2_{\rm red}$/d.o.f.=1.34/314. The observed 2-10~keV X-ray flux was 
5.7$\times$10$^{-12}$~erg/cm$^2$/s. In order to improve the fit, we considered first a bulk motion Comptonization model \citep[the BMC model accounts for the soft thermal emission from the neutron star and its 
Comptonization in a self-consistent way;][]{titarchuk97}, and then tried a model comprising a {\sc mekal} (i.e. emission from a thin thermal medium) and a 
power-law component \citep[see discussion in][]{bozzo10}. 
The BMC model provided a statistically acceptable fit (see Table~\ref{tab:fits}), but only an upper limit could be obtained on the 
spectral index parameter $\alpha$=$\Gamma$-1. This is due to the limited energy coverage of the \xmm\ spectra (0.3-12~keV). 
Other spectral models with similar number of free parameters, including a cut-off power-law or a power-law with an exponential roll-over, provided also reasonably good 
fits to the data. However these models gave very low value of the cut-off and exponential roll-over energies ($\sim$4~keV) and unlikely negative power-law photon indices 
($\Gamma$$\sim$-0.5). We checked {\it a posteriori} that different choices of the energy binning and background extraction region would not affect these results. 
 \begin{table*} 	
 \begin{center} 	
 \tiny
 \caption{Simultaneous fit to the EPIC-pn, MOS1 and MOS2 spectra of the entire observation and the time intervals displayed in the bottom panel of Fig.~\ref{fig:param}.} 	
 \label{tab:fits} 	
 \begin{tabular}{llllllllllll} 
 \hline 
 \hline 
 \noalign{\smallskip}  
 Model   & $N_{\rm H}$  & $\Gamma$  & $kT$ & $N_{\rm mekal} $ & $\log{A}$ & $N_{\rm BMC}$ & $F_{\rm X}$ & $C_{\rm MOS1}$ & $C_{\rm MOS2}$ & $\chi^2_{\rm red}$/dof  \\ 
 \noalign{\smallskip} 
 \hline
  \noalign{\smallskip} 
   Tot. \\ 
  \hline 
 \noalign{\smallskip} 
ph*BMC	&	3.57$\pm$0.17 & $<$2.5 & 1.45$_{-0.07}^{+0.06}$ & --- & 2.17$_{-0.12}^{+0.10}$ & (8.3$_{-0.7}^{+0.9}$)$\times$10$^{-3}$ & 5.6 & 1.05$\pm$0.03 & 1.10$\pm$0.02 & 1.04/312 \\
 \noalign{\smallskip} 
ph*(mkl+pl) &  	7.0$\pm$0.3 & 1.66$\pm$0.06 & 0.113$_{-0.003}^{+0.043}$ & 68.1$_{-54.4}^{+107.8}$ & --- & --- & 5.7 & 1.04$\pm$0.03 & 1.11$\pm$0.03 & 1.14/312 \\
  \noalign{\smallskip} 
 \hline 
  \noalign{\smallskip}
  \noalign{\smallskip} 
   Spectrum ``a'' \\ 
  \hline 
  \noalign{\smallskip}
ph*pl &  5.6$\pm$0.3 & 1.63$\pm$0.07 & --- & --- & --- & --- & 4.3 & 1.06$\pm$0.04 & 1.17$\pm$0.04 & 1.26/281 \\
  \noalign{\smallskip}
ph*BMC & 3.2$\pm$0.2	& $<$2.5 & 1.34$_{-0.09}^{+0.07}$ & --- & 2.15$_{-0.15}^{+0.11}$ & (5.5$_{-0.8}^{+0.2}$)$\times$10$^{-3}$ & 4.3 & 1.06$_{-0.04}^{+0.02}$ & 1.17$\pm$0.04 & 1.10/279 \\
  \noalign{\smallskip}
ph*(mkl+pl) & 6.2$\pm$0.4  & 1.73$\pm$0.08 & 0.112$_{-0.05}^{+0.050}$ & 32.5$_{-31.5}^{+64.2}$ & --- & --- & 4.3 & 1.06$\pm$0.04 & 1.18$\pm$0.04 & 1.15/279 \\ 
  \noalign{\smallskip}
 \hline 
  \noalign{\smallskip} 
  \noalign{\smallskip} 
   Spectrum ``b'' \\ 
  \hline 
 \noalign{\smallskip} 
ph*pl &  11.3$_{-1.7}^{+2.0}$ & 1.8$\pm$0.3 & --- & --- & --- & --- & 5.1 & 1.11$\pm$0.13 & 1.12$\pm$0.14 & 1.02/69 \\
  \noalign{\smallskip}
ph*BMC & 7.8$_{-1.4}^{+1.7}$ 	& 2.5 (fixed) & 1.2$\pm$0.2 & --- & 2.17 (fixed) & (9.9$_{-0.8}^{+0.9}$)$\times$10$^{-5}$ & 5.0 & 1.11$\pm$0.12 & 1.13$\pm$0.14 & 0.97/69	\\
  \noalign{\smallskip} 
 \hline 
  \noalign{\smallskip}
  \noalign{\smallskip} 
   Spectrum ``c'' \\ 
  \hline 
 \noalign{\smallskip} 
 ph*pl & 7.9$_{-3.5}^{+3.9}$ & 1.1$\pm$0.5 & --- & --- & --- & --- & 21.7 & 1.0$\pm$0.2 & 1.0$\pm$0.2 & 0.76/31 \\
  \noalign{\smallskip}
ph*BMC & 4.6$_{-2.8}^{+3.4}$ 	& 2.5 (fixed) & 1.8$_{-0.7}^{+0.5}$ & --- & 2.17 (fixed) & (4.7$_{-1.0}^{+2.1}$)$\times$10$^{-4}$ & 21.2 & 1.0$\pm$0.2 & 1.0$\pm$0.2 & 0.75/31	\\
  \noalign{\smallskip} 
 \hline 
  \noalign{\smallskip}
  \noalign{\smallskip}  
   Spectrum ``d'' \\ 
  \hline 
 \noalign{\smallskip} 
 ph*pl & 6.5$_{-1.1}^{+1.3}$ & 1.1$\pm$0.2 & --- & --- & --- & --- & 30.7 & 1.1$\pm$0.1 & 1.1$\pm$0.1 & 1.00/90 \\
  \noalign{\smallskip}
ph*BMC & 4.1$_{-0.8}^{+1.0}$ 	& 2.5 (fixed) & 1.7$_{-0.2}^{+0.3}$ & --- & 2.17 (fixed) & (6.4$_{-0.8}^{+1.0}$)$\times$10$^{-4}$ & 30.0 & 1.1$\pm$0.1 & 1.1$\pm$0.1 & 0.95/90	\\
  \noalign{\smallskip}
 \hline 
  \noalign{\smallskip}
  \noalign{\smallskip} 
   Spectrum ``e'' \\ 
  \hline 
 \noalign{\smallskip}
 ph*pl & 7.8$\pm$1.0 & 1.3$\pm$0.2 & --- & --- & --- & --- & 13.5 & 1.1$\pm$0.1 & 1.1$\pm$0.1 & 0.89/105 \\
  \noalign{\smallskip}
ph*BMC & 5.2$_{-0.7}^{+0.8}$ 	& 2.5 (fixed) & 1.5$\pm$0.2 & --- & 2.17 (fixed) & (2.7$_{-0.2}^{+0.3}$)$\times$10$^{-4}$ & 13.1 & 1.1$\pm$0.1 & 1.1$\pm$0.1 & 0.89/105	\\
  \noalign{\smallskip} 
  \hline 
  \noalign{\smallskip} 
  \noalign{\smallskip} 
   Spectrum ``f'' \\ 
  \hline 
 \noalign{\smallskip} 
ph*pl & 7.1$\pm$0.5 & 1.8$\pm$0.1 & --- & --- & --- & --- & 5.7 & 1.05$\pm$0.06 & 1.12$\pm$0.07 & 1.16/189 \\
  \noalign{\smallskip}
ph*BMC & 4.5$_{-0.4}^{+0.5}$ 	& 2.5 (fixed) & 1.15$\pm$0.08 & --- & 2.17 (fixed) & (9.5$\pm$0.4)$\times$10$^{-5}$ & 5.7 & 1.05$\pm$0.06 & 1.12$\pm$0.07 & 1.12/189	\\
  \noalign{\smallskip}
  \hline
  \hline
\end{tabular}
\end{center} 
\tablefoot{We indicated above with ``ph'', ``mkl'', ``BB'', and ``pl'' the {\sc phabs}, {\sc mekal}, {\sc blackbody}, and the power-law component of the spectral 
model used in {\sc xspec}. In the BMC model $\alpha$=$\Gamma$-1 is the energy spectral index, $\log{A}$ gives a measurement of the fraction 
of the Compton-reprocessed emission with respect to the directly observed black-body emission, and $N_{\rm BMC}$ can be used to estimate the size of the thermally emitting 
region (see also Sect.~\ref{sec:discussion}). We did not report explicitly the normalization of the power-law component in the ph*pl, as this parameter would 
not provide additional information beside the source flux already given in the fourth column on the right). 
$C_{\rm MOS1}$ and $C_{\rm MOS2}$ are the normalization constants introduced in the model to account for inter-calibration uncertainties between 
the EPIC cameras ($C$=1 for the EPIC-pn). The absorption column density reported here is given in units of 10$^{22}$~cm$^{-2}$, $kT$ in keV, and the flux 
in units of 10$^{-12}$~erg/cm$^2$/s (2-10~keV energy range, not corrected for absorption). } 
\end{table*} 
\begin{figure}
\centering
\includegraphics[scale=0.35,angle=-90]{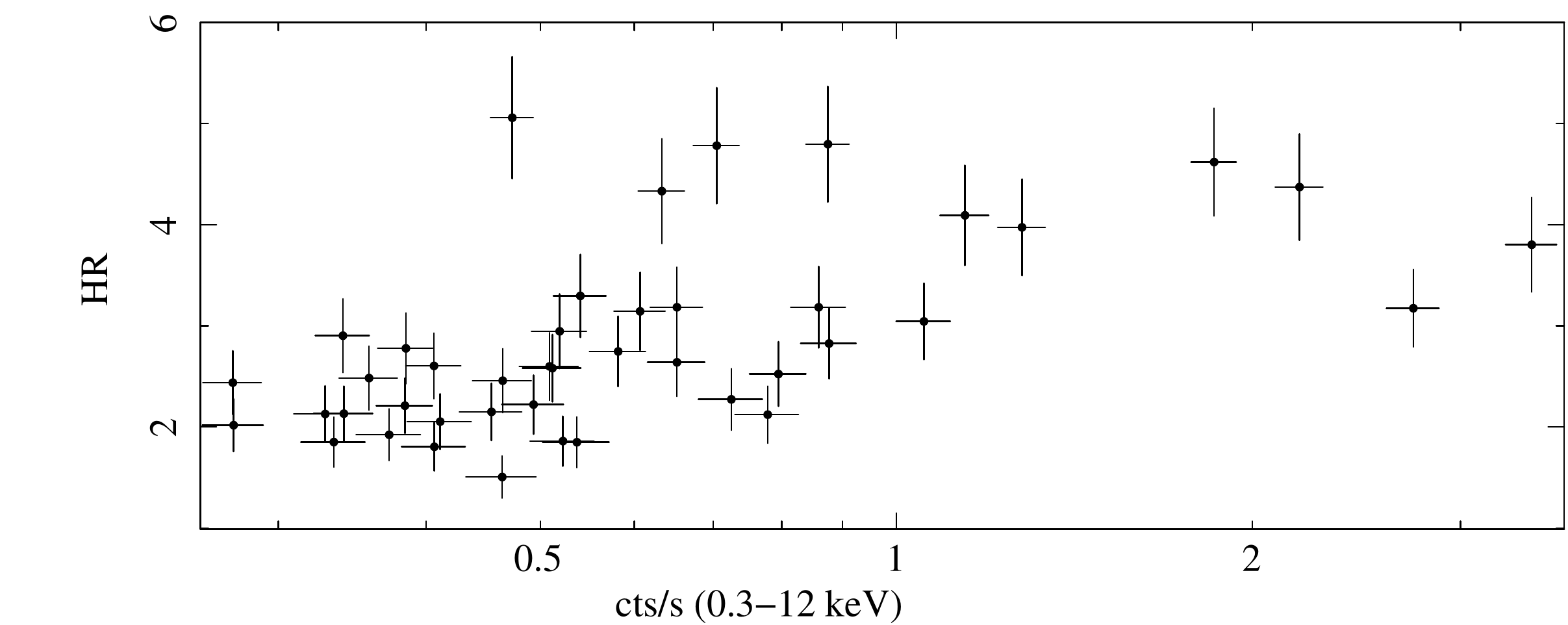}
\caption{HID of the source realized by using the same lightcurves shown in Fig.~\ref{fig:lcurve} adaptively rebinned to a 
S/N=9 (see also Fig.~\ref{fig:param}).}    
\label{fig:hr} 
\end{figure}

\begin{figure}
\centering
\includegraphics[scale=0.6,angle=-90]{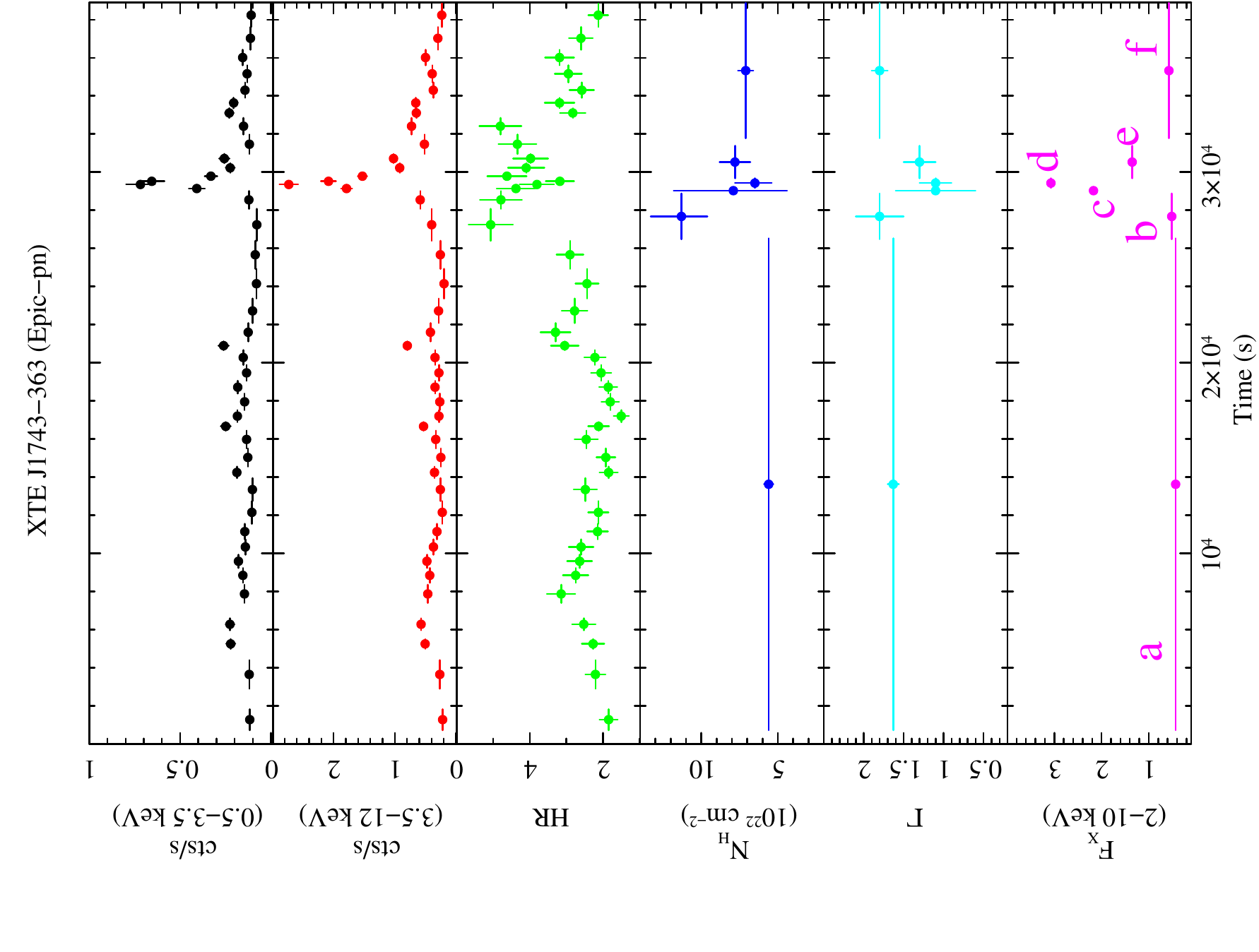}
\caption{The first two panels from the top show the source lightcurves of Fig.~\ref{fig:lcurve} rebinned adaptively to a S/N=9 
(based on the lightcurve in the 0.3-3.5 keV energy band). The third panel from the top shows the HR, calculated as the ratio between 
the source count-rate in the 3.5-12~keV vs. count-rate in the 0.3-3.5~keV energy band. The other three panels show the absorption column density 
($N_{\rm H}$), the power-law photon index ($\Gamma$), and the observed flux (units of 10$^{-11}$~erg/cm$^2$/s) in different time intervals as revealed from 
our time resolved spectral analysis (see text for details).}     
\label{fig:param} 
\end{figure}

In order to study possible spectral variations during the observation, we extracted the hardness-intensity diagram (HID) 
of the source \citep[this was calculated as in our previous papers, see e.g.][]{bozzo10}. 
From Figure~\ref{fig:hr} we noticed that the hardness ratio (HR) could not be easily 
related to changes of the source overall intensity.  
To better investigate the behavior of the HR, we rebinned the source lightcurve adaptively in order to have in each bin a signal-to-noise 
ratio of S/N=9. This is shown in Fig.~\ref{fig:param}.  
\begin{figure}
\centering
\includegraphics[scale=0.35]{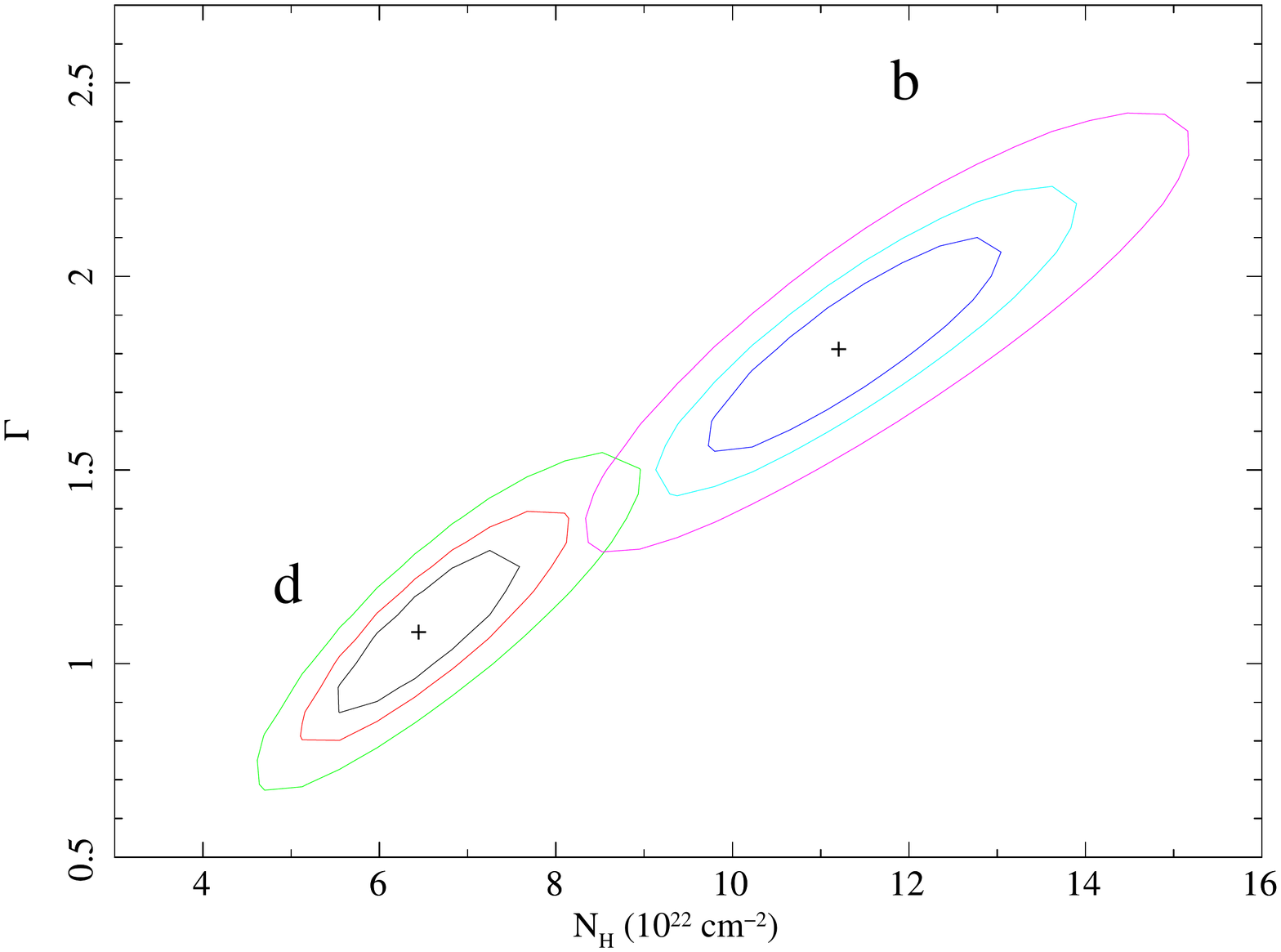}
\includegraphics[scale=0.35]{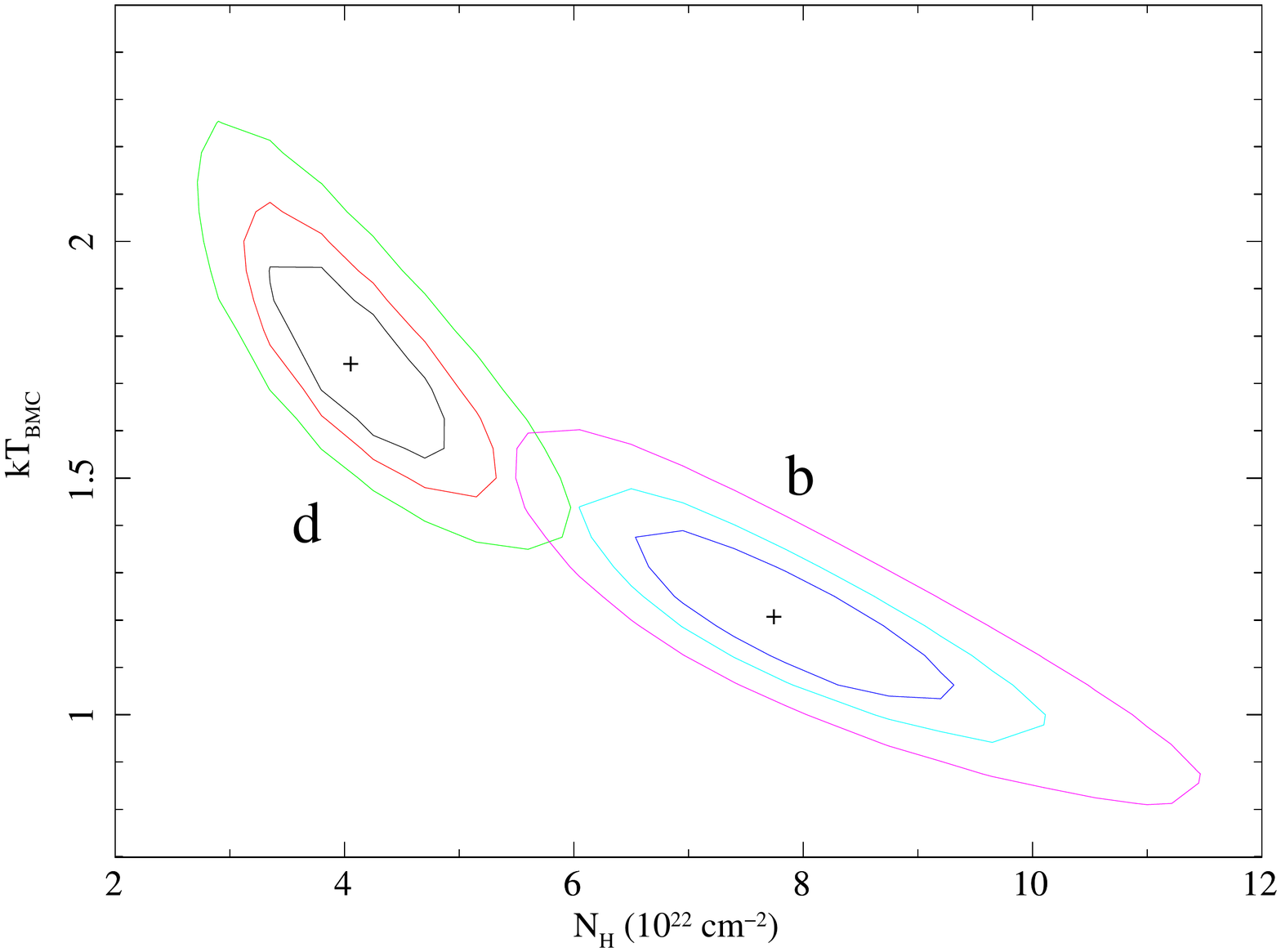}
\caption{Contours plot (68\%, 90\%, and 99\% c.l.) obtained from the fit to the spectra ``b'' and ``d'' with the ph*pl and ph*BMC models (see Fig.~\ref{fig:param} 
and Table~\ref{tab:fits}).}     
\label{fig:contours} 
\end{figure}
The complex behavior of the source HR is especially puzzling shortly before until after the bright flare occurring at $t$$\sim$30~ks from 
the beginning of the observation. In particular, it is evident that the HR significantly increases about 3~ks before the peak of the flare, displays a sudden 
drop of few hundred seconds immediately after the peak, and then decreases back to the value of the first 
part of the observation (showing similar but less pronounced variations). 
We investigated the origin of this variability by dividing the total 
observation in 6 time intervals (see Fig.~\ref{fig:param}) and performing a time resolved spectral analysis.  
EPIC-pn and EPIC-MOS spectra extracted during the same time intervals were fit together to increase the statistics and thus reduce 
the uncertainties in the fit parameters. 
Given the shorter exposure times and lower statistics with respect to the spectrum of the entire observation, all time-resolved 
spectra could be well fit by using a simple power-law model (the only exception being the spectrum ``a'', which has 
the longest exposure time). For consistency with the results obtained from the total spectrum, we also fit the time resolved spectra 
with the BMC model (we fixed in most cases the value of $log(A)$ and $\alpha$, as these parameters could not be constrained). 
The results are given in Table~\ref{tab:fits} and shown in Fig.~\ref{fig:param}.  
The fits to the time-resolved spectra with all models suggest there was a significant change in the spectral parameters 
during the bright flare. The largest variation occurs between the time intervals ``b'' and ``d''.  
To demonstrate that these changes are not model dependent, we also show in Fig.~\ref{fig:contours} the contour plots of the  
spectral parameters of the absorbed BMC and the power-law model that displayed a significant change during these intervals. 
Note that the variation of the $N_{\rm H}$ revealed from the fit with a simple absorbed power-law model is fully consistent with 
the ph*(mkl+pl) interpretation of the X-ray spectrum. During the time passed from the ``b'' to the ``d'' time intervals ($\lesssim$1.5~ks) 
we would not expect significant changes in the {\sc mekal} component, as the latter should not vary on time-scales shorter than 
$R_{\rm mekal}$/$c$$\sim$2.3~ks (here $c$ is the speed of light and $R_{\rm mekal}$$>$10$^{13}$~cm as estimated in Sect.~\ref{sec:discussion}). 
We verified that the addition of a {\sc mekal} component with temperature and normalization fixed to the values determined from the total spectrum to the time 
resolved spectra would not significantly affect the results presented in Fig.~\ref{fig:contours}. 

The results in Fig.~\ref{fig:param} and \ref{fig:contours} suggest that the absorption column density 
in the direction of the source raised up about 3~ks before the onset of the flare, dropped close to the peak of the event (when the source reached 
the highest X-ray flux), and then raised again during the initial decay of the flare. Correspondingly, we observed a flattening of the power-law 
photon index (in the fit with the model ph*pl) or an increase in the temperature of the seed thermal photons in the fit with the ph*BMC model.  
The interpretation of these results is discussed in Sect.~\ref{sec:discussion}. 
We note that fixing in the fit to the spectrum ``b'' the same value of the absorption column density measured for spectrum 
``d'' would give unacceptable results in both cases of the ph*pl and ph*BMC models ($\chi^2_{\rm red}$/d.o.f.=1.41/70 and 1.37/70, respectively). 
The same conclusion applies if the absorption column density measured for spectrum ``b'' is used and fixed in the fit to the spectrum ``d''. 
In this case we obtained for the ph*pl and ph*BMC models $\chi^2_{\rm red}$/d.o.f.=1.30/91 and 1.23/91, respectively.  

\begin{figure}
\centering
\includegraphics[scale=0.37,angle=270]{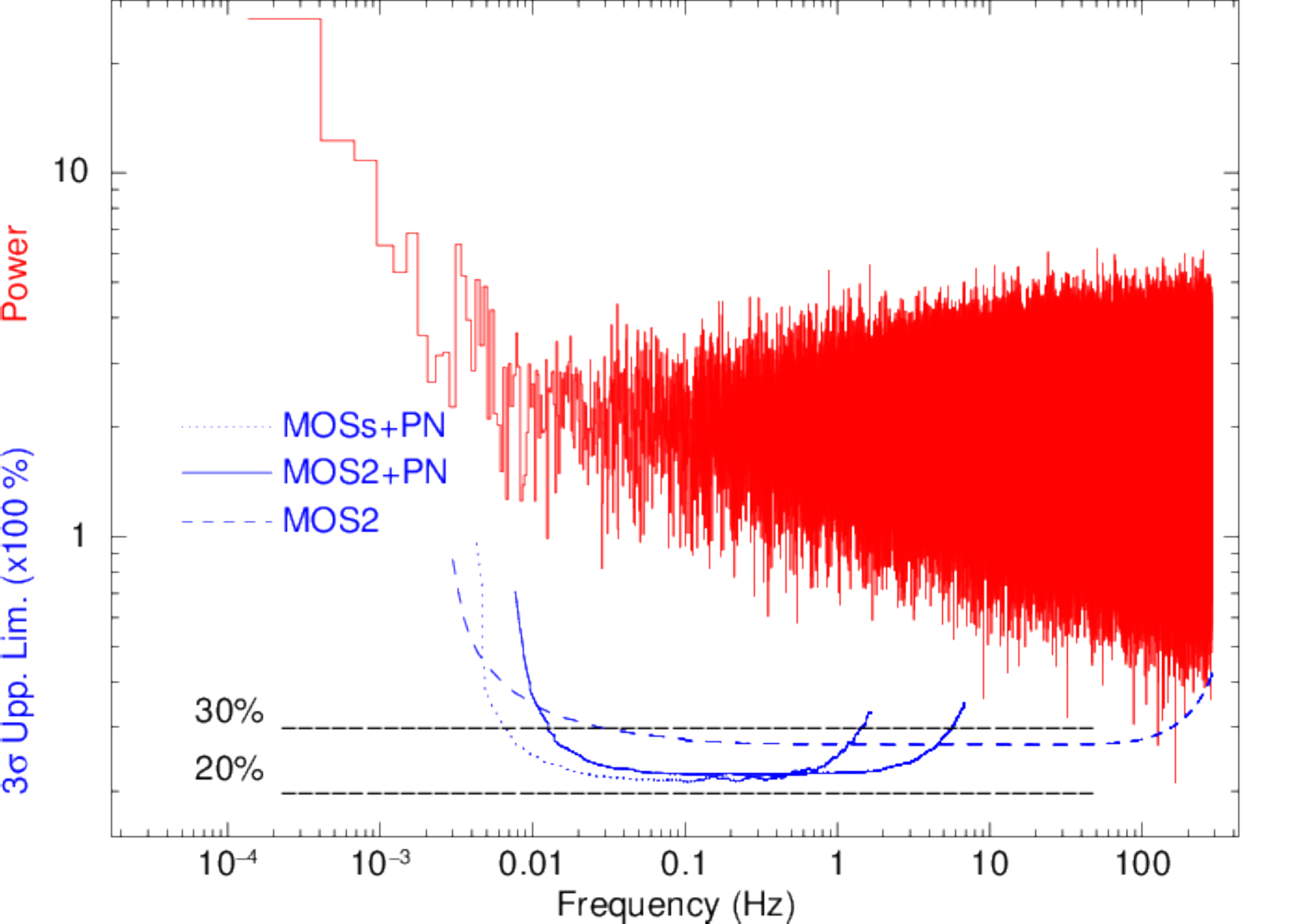}
\caption{In the upper part of the plot we show the power spectrum produced using the MOS2 event list as an example. 
In the bottom part of the plot we show the curves representing the upper limits to the non-detection of pulsations 
calculated according to \citet{israel96}. The solid, dashed and dotted lines correspond to the different sampling times and event list combinations 
discussed in the text. The two horizontal dashed lines represent the 20\% and 30\% upper limit levels on the pulsed fraction upper limits as a function of the 
frequency. The most stringent upper limits we could provide are at 20–30\% for frequencies in the range 0.005–200 Hz.}     
\label{fig:uls} 
\end{figure}

The source and background EPIC MOS and pn event lists were used to carry out an in-depth search for coherent 
signals after barycenter correction. We applied to the lists of barycentered photon arrival times 
the power-spectrum search algorithm developed by \citet{israel96}. This method is optimized 
to search for periodicities in ``coloured'' power spectrum components and provides upper limits if no signal is detected. 
Due to the different sampling times  of the three EPIC cameras we decided to carry out the search in three different ways:  
(i) by using all the event lists and a binning time of 0.3s, (ii) by combining the MOS2 and pn event lists with binning time of 73.4ms (keeping the original 
Fourier resolution), and by using only the MOS2 event list with a binning time of 1.75ms (thus maximizing the Nyquist frequency). 
No significant signal was found. We show in Fig.~\ref{fig:uls} the derived 3$\sigma$ upper limits to the pulsed fraction as a function of 
the pulse frequency. 

\section{ \swift\ data analysis and results}
\label{sec:swift}

\swift\,/XRT data were obtained as a target of opportunity (ToO)
monitoring campaign. The ToO started on 
2012 April 19 with 1\,ks per day until July 20; then 1\,ks observations were carried out every other day.  
The campaign lasted 115 days, with 82 observations 
for a total on-source exposure of 79\,ks.  
We also considered the archival \swift\ observation 00037884001 from February 2010 (see Table~\ref{tab:xrtdata} for details). 

The X-ray Telescope \citep[XRT, ][]{burrows05} data were processed with 
standard procedures ({\sc xrtpipeline} v0.12.6), filtering and screening criteria were applied 
by using {\sc ftools} in the {\sc heasoft} package (v.6.12). 
Given the low count rate of the source throughout the monitoring campaign,  
we only considered photon-counting (PC) mode data, 
and selected event grades 0--12 (\citealt{burrows05}). 
We used the latest spectral redistribution matrices in CALDB (20120713).  
The best source position determined with XRT is at $\alpha_{\rm J2000}$=17$^{\rm h}$43$^{\rm m}$01$^{\rm s}$.31 and 
$\delta_{\rm J2000}$=-36$^{\circ}$22$'$21$''$.4, with an associated uncertainty of 2.3$''$ at 90\% c.l.\footnote{We used the on-line tool 
http://www.swift.ac.uk/user\_objects/; see \citet{evans09}.} 

\begin{figure}
\centering
\includegraphics[scale=0.35, angle=270]{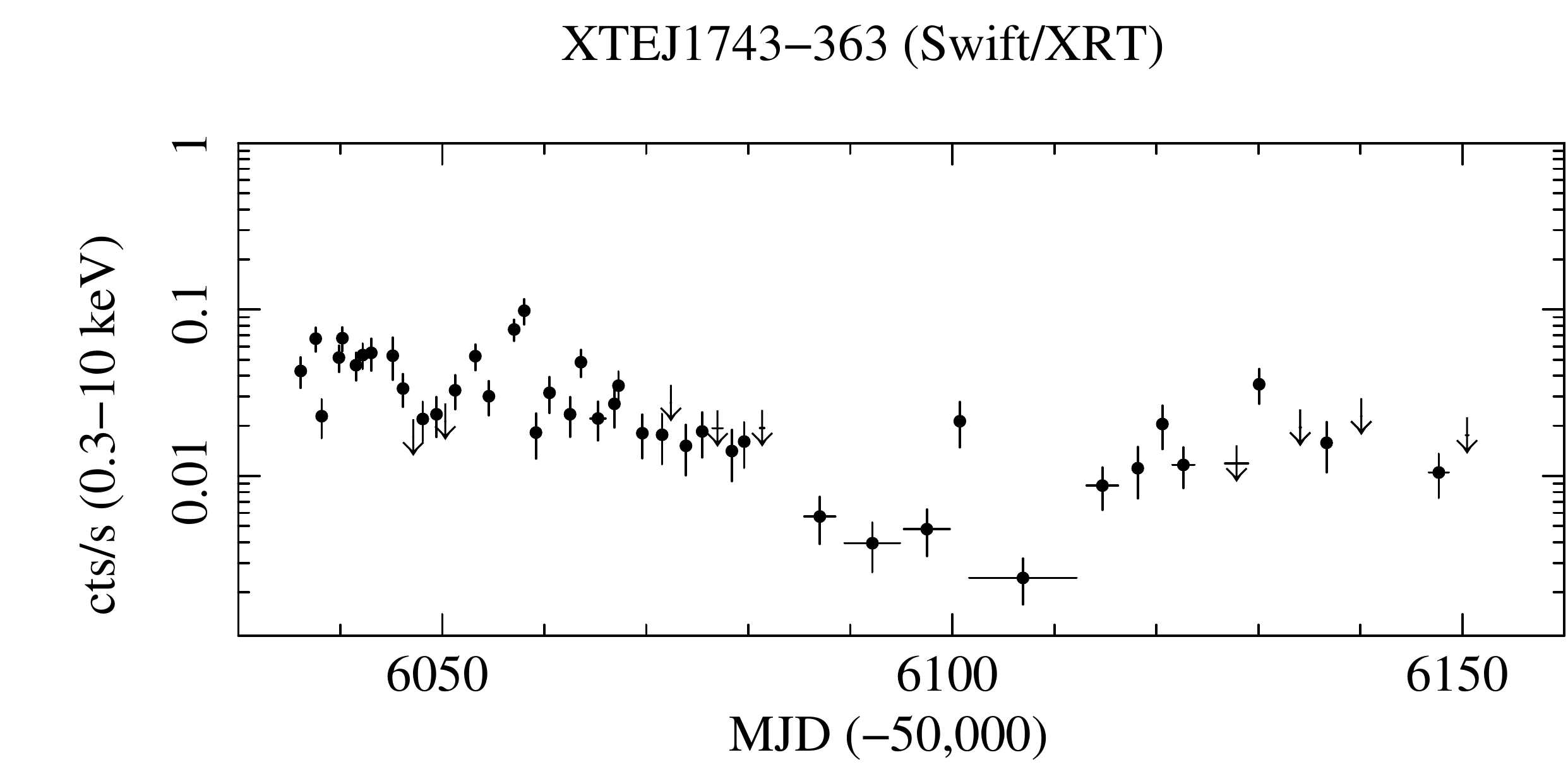}
\caption{Long-term monitoring of \xte\ performed with \swift\,/XRT (the campaign started about 50 days after the \xmm\ observation). The count-rate 
is given in the 0.3-10 keV energy range. 
Assuming the source averaged spectral 
properties reported in Sect.~\ref{sec:swift}, a count-rate of 0.01~cts/s corresponds to about 1.2$\times$10$^{-12}$~erg/cm$^2$/s. Downward arrows 
correspond to 3$\sigma$ upper-limits on the source count-rate.}    
\label{fig:xrtlcv} 
\end{figure}

The 0.3--10\,keV XRT background-subtracted lightcurve was created at a 1~day resolution  
and shown in Fig.~\ref{fig:xrtlcv}. All count-rate measurements have been corrected for PSF losses and vignetting. 
The source could not be detected in all observations. 
In order to preserve a good sampling in time of the source count-rate, we stacked together close-by observations 
in which a detection of the source could not be obtained after a few days of integration ($\sim$2-3). For all 
isolated non-detections, we indicated in the figure the corresponding 3$\sigma$ upper-limits on the source count-rate 
with downward arrows. We did not include in Fig.~\ref{fig:xrtlcv} the observation 00037884001. At the epoch of this observation, 
XRT did not detect the source and we derived a 90\% c.l. upper limit on its X-ray flux of 8$\times$10$^{-13}$~erg/cm$^2$/s (2-10~keV 
not corrected for absorption). This is the only observation in the soft X-ray domain that overlap with previously reported 
\rxte\ data (see Sect.~\ref{sec:discussion}). 

\begin{figure}
\centering
\includegraphics[scale=0.35, angle=-90]{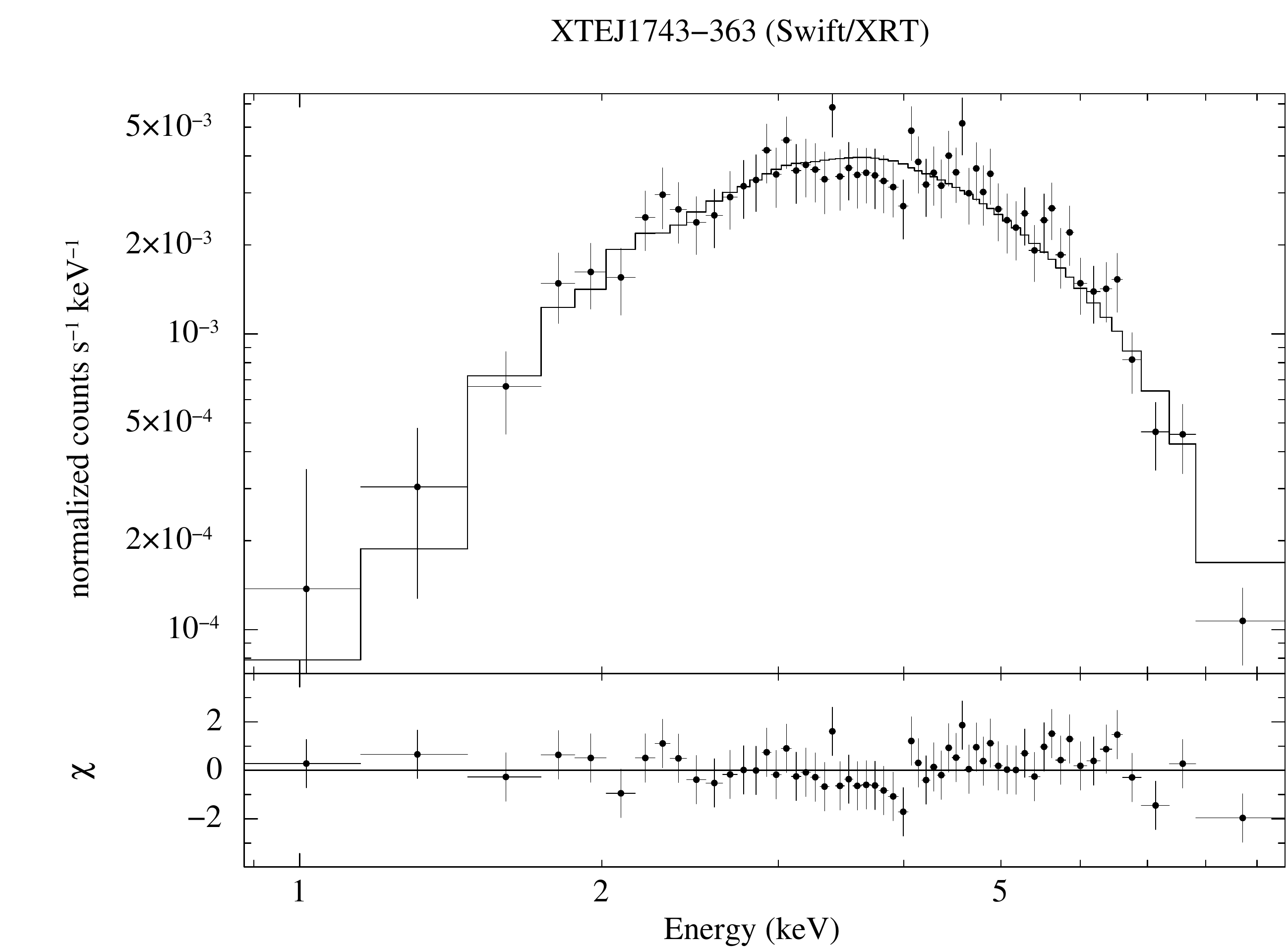}
\caption{ \swift\,/XRT spectrum extracted by summing-up all data in Table~\ref{tab:xrtdata}. The best fit is obtained 
with a simple absorbed power-law model (see text for details). The residuals from this fit are shown in the bottom panel.}    
\label{fig:swift_spe} 
\end{figure}

For our spectral analysis, we extracted the mean spectrum in the same regions as 
those adopted for the light curves; ancillary response files were 
generated with {\sc xrtmkarf} to account for different extraction regions, 
vignetting, and PSF corrections. 
The data were rebinned with a minimum of 20 counts per energy bin 
and were fit  in the 0.3--10\,keV energy range. 
The XRT spectrum (see Fig.~\ref{fig:swift_spe}) could be well fit ($\chi^2_{\rm red}$/d.o.f.=0.90/56) with a simple absorbed power-law 
model. We obtained $N_{\rm H}$=(5.6$\pm$0.9)$\times$10$^{22}$~cm$^{-2}$ and $\Gamma$=1.6$\pm$0.3, in reasonably  
good agreement with the average results obtained from \xmm\ (Sect.~\ref{sec:data}). The averaged observed 2-10~keV X-ray flux 
was 2.0$\times$10$^{-12}$~erg/cm$^2$/s. For comparison with results in Sect.~\ref{sec:data} we also fit the XRT spectrum with the ph*BMC model 
($\chi^2_{\rm red}$/d.o.f.=0.65/56). In this case we obtained $N_{\rm H}$=(3.4$_{-0.6}^{+0.8}$)$\times$10$^{22}$~cm$^{-2}$, $kT$=1.1$\pm$0.2~keV, 
and $N_{\rm BMC}$=(2.8$\pm$0.3)$\times$10$^{-5}$. As XRT recorded a total of only 1244 counts from the source, we did not attempt a time 
resolved spectral analysis of the XRT data.  

\section{Discussion and conclusions}
\label{sec:discussion}

In this paper, we reported for the first time an in-depth monitoring of \xte\ in the soft X-ray domain (0.3-12~keV). 
The position of the source determined from these data at a few arcsec accuracy is compatible with that suggested previously 
by \citet{ratti10}, thus confirming the association of \xte\ with the M8III giant star identified by \citet{smith12} and the 
classification of the source as a SyXBs. 

In the \xmm\ data, the source displayed variability on time-scales of few hundred to thousand of seconds. 
A relatively bright flare occurred about 30~ks after the beginning of the observation and lasted for a few ks. 
This behavior is reminiscent of what is often observed in neutron star (NS) wind accreting binaries. 
This conclusion is also supported by the spectral analysis. The average \xmm\ spectrum could be described reasonably well with 
models usually adopted for wind accreting systems. The BMC model provided the best description of the data, but reasonably good 
results could also be obtained by using a model comprising a thin thermal emission component and a power-law. 
The BMC model describes with a self-consistent approach the case in which blackbody seed photons from the NS are subjected to bulk and thermal Comptonization 
in an optically thin regime \citep{titarchuk97}. In this model the parameter $log(A)$ gives an indication of the fraction of the Comptonized 
thermal photons with respect to those that are directly visible (for $log(A)$=8 there is no thermal emission directly visible, whereas for 
$log(A)$=-8 there is no effect due to the Comptonization). From the normalization of the BMC model it is also possible to estimate the extension of the 
thermal emitting region according to the equation \citep[see e.g.,][and references therein]{bozzo10}
\begin{equation}
R_{\rm BMC} = \frac{88}{C^{1/2}}\frac{N_{\rm BMC}^{1/2} d_{\rm 10 kpc}}{(kT)_{\rm 1 keV}^2}\,{\rm km}
\label{eq:mekal} 
\end{equation} 
where $(kT)_{\rm 1~keV}$ is the temperature of the thermal emitting component in units of 1~keV, and C=0.25  
for an emitting surface with the geometry of a circular slab. Equation \ref{eq:mekal} and 
the results of the fit to the total spectrum in Table~\ref{tab:fits}  (where the spectral parameters are better constrained) suggest a radius for the thermal 
emitting region of 1 to a few km \citep[for a distance of $\sim$5~kpc;][]{smith12}. Such an extended ``hot-spot'' would 
be expected in case of wind-fed systems accreting at low luminosities \citep[see discussion in][and references therein]{bozzo10}. We note that at 5~kpc 
the peak luminosity recorded by \xmm\ from \xte\ is $\sim$10$^{35}$~erg/s. As the temperature of the 
emitted thermal radiation is about 1-2~keV and the energy coverage of the \xmm\ data is limited to 12~keV, we could not well constraint the 
spectral index $\alpha$.  

Even though the BMC model was statistically preferable, a relatively good description of the data could also be obtained by using a model comprising a 
power-law and a {\sc mekal} component. The latter component accounts in this model for the excess that emerge at energies 
$\lesssim$2~keV when a simple absorbed power-law model is used to fit the data (see Fig.~\ref{fig:totalspectrum}). Similar ``soft excesses'' have been found in a number of 
SFXT sources \citep{bozzo10,sidoli10}, SyXBs \citep{masetti06}, and is thought to be an ubiquitous characteristics of all NS accreting X-ray binaries \citep{hickox04}. 
As discussed in \citet{bozzo10}, the {\sc mekal} component in wind-fed systems with accreting NSs might represent the contribution to the total X-ray 
emission from the wind of the companion star. The normalization of the {\sc mekal} component can also be translated into an estimate of the size of the emitting 
region \citep[see, e.g.,][and reference therein]{bozzo10}
\begin{equation}
R_{\rm mekal}= \sqrt[3]{\frac{3 N_{\rm mekal}}{10^{-14}}\left(\frac{d}{n_{\rm H}}\right)^2} \simeq 7\times10^{13} d_{\rm 5 kpc}^{2/3} a_{\rm 13}^{2/3}~{\rm cm}, 
\end{equation} 
where $d_{\rm 5kpc}$ is the source distance in units of 5~kpc, $n_{\rm H}$$\sim$$N_{\rm H}$/$a$, $a_{\rm 13}$ is the binary separation in units of 
10$^{13}$~cm and we made use of the results in Table~\ref{tab:fits} for $N_{\rm mekal}$ and $N_{\rm H}$.   
This is consistent with the soft X-ray emission in \xte\ being produced in the surrounding of a NS hosted in a wide binary system. As the orbital period of \xte\ is not 
known, it is not presently possible to assess the applicability of this model. We remark that a similar soft spectral component was 
detected during a \beppo\ observation of the SyXB 4U\,1954+31 and interpreted in the same way \citep{masetti06}. 

In Sect.~\ref{sec:data} we analyzed in detail the behavior of the HR recorded during the \xmm\ observation.  We showed in particular 
that the HR underwent a clear change shortly before until after the bright flare detected about 30~ks after the beginning of the observation. 
Our analysis showed that there was an increase of the  absorption column density 
before the onset of the event followed by a sudden drop at the peak of the flare for about few hundreds seconds. 
This behavior of the HR is similar to what was observed during a flare from the SFXT IGR\,J18410-0535 caught by \xmm\ in 2011 \citep{bozzo11}.
In that occasion the flare was ascribed to an episode of enhanced accretion onto the NS 
due to the encounter with a ``clump'' of material from the stellar wind. The drop in the absorption column density at the peak of the flare was interpreted as being 
due to the photo-ionization of the clump by X-rays from the NS. 
If a similar interpretation is applied to the flare from \xte,\ the calculations in \citet{bozzo11} suggest in this case a clump radius 
of $R_{\rm cl}$$\sim$2.2$\times$10$^{10}$$v_{\rm w7}$~cm, where we scaled the wind velocity to a value that is more appropriate for a M giant star 
\citep[$\sim$50-500~km/s; see, e.g.,][and references therein]{espey08,lu12}. At this low velocity, the accretion radius of the NS is   
$R_{\rm acc}$=2$G$$M$/$v_{\rm w}$$^2$=3.7$\times$10$^{12}$$v_{\rm w7}$$^{-2}$~cm, and thus becomes larger than the estimated size of the 
clump\footnote{We neglected in the calculation of $R_{\rm acc}$ the contribution of the NS orbital velocity \citep[see, e.g.,][]{bozzo08}. For the range of system parameters 
(masses, orbital periods and wind velocities) considered here for a SyXB, the NS orbital velocity would be comparable or lower than $v_{\rm w}$ and thus this 
approximation does not significantly affect our conclusions.}.    
At variance with the case of IGR\,J18410-0535, we thus can not make use here of the equation $M_{\rm cl}$=$M_{\rm acc}$($R_{\rm cl}$/$R_{\rm acc}$)$^2$ from 
\citet{bozzo11}, but we have to infer the mass of the clump directly from the observation, i.e. $M_{\rm cl}$$\simeq$$M_{\rm acc}$. 
The latter can be estimated as $M_{\rm acc}$=9$\times$10$^{17}$$d_{\rm 5~kpc}^2$~g by using the observed flare duration of 4.4 ks and integrating over time the 
unabsorbed flux measured during the intervals b, c, d, and e in Fig.~\ref{fig:param}. The absorption column density caused by the passage of this clump along the line of sight to the X-ray source is 
$N_{\rm H}$$\sim$$M_{\rm cl}$/($R_{\rm cl}^2$$m_{\rm p}$)$\sim$10$^{21}$$v_{w7}^{-2}$$d_{\rm 5~kpc}^2$~cm$^{-2}$, thus suggesting for \xte\ a particularly low velocity  
(the absorption column density measured from the \xmm\ spectra is $N_{\rm H}$$\sim$10$^{23}$~cm$^{-2}$, see Table~\ref{tab:fits}). 
We note that complex accretion environments and inhomogeneous winds in SyXBs were already suggested in the case of 4U\,1954+31, which X-ray spectrum 
revealed the presence of multiple and variable absorbers close to the NS \citep{mattana06,masetti06}. 
As mentioned also for IGR\,J18410-0535 and other wind accreting binaries, the NS magnetic field and rotation might also play a role regulating the amount of material 
that can be accumulated and accreted close to the star magnetospheric boundary \citep{grebenev07,bozzo08,postnov11}. 
The ``magnetic and centrifugal gates'' depends on the local condition at the NS magnetospheric boundary, and in particular on the magnetic field strength and 
the rotation velocity of the compact object. The lack of information on the properties of the NS hosted in \xte\ and the system geometry (orbital period, 
eccentricity, ...) prevents a detailed investigation on how these gates can affect the parameters of the clump reported above. 
Further observations of SyXBs with X-ray instruments endowed with a large collecting area in the soft X-ray domain can help investigating their flaring behavior in more 
details and understanding the properties of winds from cool stars. The candidate ESA mission LOFT \citep{feroci12} can provide significant better capabilities 
in these respects compared to the present generation of X-ray telescopes. The unprecedented large effective area of its on-board Large Area Detector (LAD), 
operating in the energy range 2-30~keV, will be able to detect spectral variations during enhanced X-ray emission episodes in bright accreting binaries down to time 
scales of few seconds \citep[see also discussion in][]{bozzo12b}.   

Our long term monitoring campaign with XRT (see Sect.~\ref{sec:swift}) evidenced a 40~days-long period of particularly puzzling low X-ray intensity 
of the source centered around 56100~MJD (see Fig.~\ref{fig:xrtlcv}).  This period 
was interrupted only by a single flare occurring at $\sim$56100~MJD (in the corresponding XRT pointing the source displayed a  
count-rate of $\sim$2$\times$10$^{-2}$ cts/s compared to an average count-rate of $\sim$5$\times$10$^{-3}$ cts/s in the near-by observations). 
We note that similar low intensity episodes were also reported previously by \citet[][e.g. around 54115~MJD]{smith12}. In some of the \rxte\ observations 
the emission recorded by the PCA from the direction of the source was compatible with being due only to Galactic diffuse emission, suggesting that in 
these periods \xte\ was in a quiescent state ($\ll$10$^{-11}$~erg/cm$^2$/s). The latest available PCA observations performed toward the end  
of the \rxte\ campaign (from $\sim$55200~MJD to $\sim$55300~MJD) also caught \xte\ again in a low quiescent state 
\citep[see Fig.~7 in][]{smith12}. The \swift\ pointing ID.~00037884001 that was carried out in the direction of the source 
at a similar epoch (55237.82~MJD, see Sect.~\ref{sec:swift}) did not detect any significant X-ray emission and placed a tight upper limit on its 
flux at 8$\times$10$^{-13}$~erg/cm$^2$/s (2-10~keV). This is compatible with 
the flux measured by \swift\ during the low intensity episode that occurred around 56100~MJD. 
Outside these two quiescent periods, the fluxes measured by \xmm\ and \swift\ 
are a factor of 5-10 lower than that measured during the latest \rxte\ observations that still caught \xte\ undergoing some residual X-ray activity 
($\sim$3$\times$10$^{-11}$~erg/cm$^2$/s) around 54304~MJD. We thus conclude that, beside the overall decreasing trend in the source X-ray emission 
over the past 15~yrs \citep{smith12}, \xte\ also displays quiescent periods lasting several to tens of days.

\section{Acknowledgments}

We thank the Anonymous Referee for his/her constructive suggestions that helped us improving the paper. 
PR acknowledges financial contribution from the contract ASI-INAF I/004/11/0. EB acknowledges support from ISSI 
through funding for the International Team on ``Unified View of Stellar Winds in Massive X-ray Binaries'' (leader: S. Mart\'inez-Nu$\tilde{\rm n}$ez). 

\bibliographystyle{aa}
\bibliography{J1743}

\Online

 \begin{table*} 	
 \begin{center} 	
 \caption{Summary of all \swift\,/XRT pointings used in this paper.} 	
 \label{tab:xrtdata} 	
 \begin{tabular}{lllll 	  } 
 \hline 
 \hline 
 \noalign{\smallskip}  
 Sequence   & Obs/Mode  & Start time  (UT)  & End time   (UT) & Exposure    \\ 
            &           & (yyyy-mm-dd hh:mm:ss)  & (yyyy-mm-dd hh:mm:ss)  &(s)        \\
 \noalign{\smallskip} 
 \hline 
 \noalign{\smallskip} 
00037884001	&	XRT/PC	&	2010-02-10 19:41:08	&	2010-02-10 23:12:56	&	2873	\\
00037884002	&	XRT/PC	&	2012-04-19 02:43:56	&	2012-04-19 02:59:57	&	938	\\
00037884004	&	XRT/PC	&	2012-04-20 14:03:38	&	2012-04-20 14:20:56	&	1018	\\
00037884005	&	XRT/PC	&	2012-04-21 02:50:05	&	2012-04-21 05:49:56	&	1151	\\
00037884006	&	XRT/PC	&	2012-04-22 20:29:17	&	2012-04-22 20:46:58	&	1051	\\
00037884007	&	XRT/PC	&	2012-04-23 04:34:20	&	2012-04-23 04:51:58	&	1038	\\
00037884008	&	XRT/PC	&	2012-04-24 12:35:18	&	2012-04-24 12:52:56	&	1056	\\
00037884009	&	XRT/PC	&	2012-04-25 03:12:18	&	2012-04-25 06:32:57	&	1096	\\
00037884010	&	XRT/PC	&	2012-04-26 00:03:19	&	2012-04-26 01:51:57	&	634	\\
00037884011	&	XRT/PC	&	2012-04-27 03:14:23	&	2012-04-27 03:30:56	&	970	\\
00037884012	&	XRT/PC	&	2012-04-28 03:19:23	&	2012-04-28 03:35:56	&	396	\\
00037884013	&	XRT/PC	&	2012-04-29 03:19:19	&	2012-04-29 03:36:57	&	1038	\\
00037884014	&	XRT/PC	&	2012-04-30 03:26:24	&	2012-04-30 03:43:57	&	1033	\\
00037884015	&	XRT/PC	&	2012-05-01 01:53:25	&	2012-05-01 02:11:58	&	1101	\\
00037884016	&	XRT/PC	&	2012-05-02 09:54:42	&	2012-05-02 10:11:58	&	1033	\\
00037884017	&	XRT/PC	&	2012-05-03 06:52:10	&	2012-05-03 07:09:56	&	1043	\\
00037884018	&	XRT/PC	&	2012-05-04 05:19:54	&	2012-05-04 07:00:56	&	1008	\\
00037884020	&	XRT/PC	&	2012-05-06 05:28:38	&	2012-05-06 05:46:56	&	1098	\\
00037884021	&	XRT/PC	&	2012-05-07 13:32:18	&	2012-05-07 13:50:57	&	1101	\\
00037884022	&	XRT/PC	&	2012-05-10 00:45:16	&	2012-05-10 01:02:57	&	1058	\\
00037884023	&	XRT/PC	&	2012-05-11 00:45:10	&	2012-05-11 01:02:56	&	1046	\\
00037884024	&	XRT/PC	&	2012-05-12 04:02:13	&	2012-05-12 04:19:56	&	1048	\\
00037884025	&	XRT/PC	&	2012-05-13 12:09:11	&	2012-05-13 12:26:56	&	1048	\\
00037884026	&	XRT/PC	&	2012-05-14 00:59:08	&	2012-05-14 01:13:11	&	832	\\
00037884027	&	XRT/PC	&	2012-05-15 12:22:16	&	2012-05-15 12:39:57	&	1038	\\
00037884028	&	XRT/PC	&	2012-05-16 13:53:07	&	2012-05-16 14:10:57	&	1048	\\
00037884029	&	XRT/PC	&	2012-05-17 12:21:48	&	2012-05-17 12:39:56	&	1078	\\
00037884030	&	XRT/PC	&	2012-05-18 23:49:33	&	2012-05-18 23:56:57	&	431	\\
00037884031	&	XRT/PC	&	2012-05-19 20:43:36	&	2012-05-19 21:00:56	&	1028	\\
00037884032	&	XRT/PC	&	2012-05-20 00:03:19	&	2012-05-20 12:59:57	&	1048	\\
00037884034	&	XRT/PC	&	2012-05-22 05:06:26	&	2012-05-22 23:59:56	&	1181	\\
00037884035	&	XRT/PC	&	2012-05-24 12:54:35	&	2012-05-24 13:09:55	&	903	\\
00037884036	&	XRT/PC	&	2012-05-25 09:37:36	&	2012-05-25 09:53:56	&	965	\\
00037884037	&	XRT/PC	&	2012-05-26 21:06:17	&	2012-05-26 21:23:58	&	1041	\\
00037884038	&	XRT/PC	&	2012-05-28 11:19:46	&	2012-05-28 11:36:57	&	1028	\\
00037884039	&	XRT/PC	&	2012-05-29 11:37:55	&	2012-05-29 11:55:56	&	1076	\\
00037884040	&	XRT/PC	&	2012-05-30 11:20:37	&	2012-05-30 11:34:57	&	860	\\
00037884041	&	XRT/PC	&	2012-05-31 00:32:41	&	2012-05-31 17:59:57	&	1206	\\
00037884042	&	XRT/PC	&	2012-06-01 05:08:34	&	2012-06-01 23:13:57	&	1276	\\
00037884043	&	XRT/PC	&	2012-06-03 03:30:45	&	2012-06-03 13:18:57	&	1437	\\
00037884044	&	XRT/PC	&	2012-06-07 11:55:00	&	2012-06-07 12:13:56	&	1116	\\
00037884045	&	XRT/PC	&	2012-06-08 10:16:52	&	2012-06-08 10:32:57	&	958	\\
00037884046	&	XRT/PC	&	2012-06-10 12:07:51	&	2012-06-10 12:23:56	&	958	\\
00037884047	&	XRT/PC	&	2012-06-11 10:33:08	&	2012-06-11 10:49:56	&	990	\\
00037884048	&	XRT/PC	&	2012-06-12 18:33:08	&	2012-06-12 18:49:56	&	35	\\
00037884049	&	XRT/PC	&	2012-06-13 07:15:36	&	2012-06-13 07:32:56	&	1018	\\
00037884050	&	XRT/PC	&	2012-06-14 06:17:52	&	2012-06-14 12:47:57	&	1261	\\
00037884051	&	XRT/PC	&	2012-06-15 09:32:58	&	2012-06-15 22:28:56	&	1048	\\
00037884052	&	XRT/PC	&	2012-06-16 18:43:15	&	2012-06-16 20:55:58	&	1389	\\
00037884053	&	XRT/PC	&	2012-06-17 05:58:52	&	2012-06-17 06:15:58	&	1018	\\
00037884054	&	XRT/PC	&	2012-06-18 09:43:05	&	2012-06-18 09:48:56	&	351	\\
00037884055	&	XRT/PC	&	2012-06-19 04:57:59	&	2012-06-19 22:27:57	&	1023	\\
00037884056	&	XRT/PC	&	2012-06-20 20:58:53	&	2012-06-20 22:43:57	&	1146	\\
00037884057	&	XRT/PC	&	2012-06-21 00:03:19	&	2012-06-21 17:59:57	&	306	\\
00037884058	&	XRT/PC	&	2012-06-22 17:31:42	&	2012-06-22 17:47:57	&	970	\\
00037884059	&	XRT/PC	&	2012-06-23 15:59:06	&	2012-06-23 16:15:56	&	1003	\\
00037884060	&	XRT/PC	&	2012-06-24 02:06:00	&	2012-06-24 21:26:57	&	1048	\\
00037884061	&	XRT/PC	&	2012-06-25 02:10:08	&	2012-06-25 08:40:56	&	1036	\\
  \noalign{\smallskip}
  \hline
  \end{tabular}
  \end{center}
  \end{table*}

\begin{table*} 	
 \begin{center} 	
 \begin{tabular}{lllll 	  }  
 \hline 
 \hline 
 \noalign{\smallskip} 
 Sequence   & Obs/Mode  & Start time  (UT)  & End time   (UT) & Exposure    \\ 
            &           & (yyyy-mm-dd hh:mm:ss)  & (yyyy-mm-dd hh:mm:ss)  &(s)        \\
 \noalign{\smallskip} 
 \hline 
 \noalign{\smallskip} 
00037884062	&	XRT/PC	&	2012-06-26 01:56:11	&	2012-06-26 02:09:56	&	817	\\
00037884063	&	XRT/PC	&	2012-06-27 21:21:33	&	2012-06-27 23:03:57	&	1116	\\
00037884064	&	XRT/PC	&	2012-06-28 19:32:31	&	2012-06-28 19:48:56	&	963	\\
00037884065	&	XRT/PC	&	2012-06-29 16:21:51	&	2012-06-29 16:37:56	&	945	\\
00037884066	&	XRT/PC	&	2012-06-30 16:26:00	&	2012-06-30 16:41:56	&	953	\\
00037884067	&	XRT/PC	&	2012-07-04 03:49:36	&	2012-07-04 04:06:54	&	1018	\\
00037884068	&	XRT/PC	&	2012-07-05 03:52:21	&	2012-07-05 04:09:48	&	990	\\
00037884069	&	XRT/PC	&	2012-07-06 02:19:29	&	2012-07-06 02:32:54	&	800	\\
00037884070	&	XRT/PC	&	2012-07-08 05:37:58	&	2012-07-08 05:57:54	&	1181	\\
00037884071	&	XRT/PC	&	2012-07-09 18:50:41	&	2012-07-09 19:05:56	&	898	\\
00037884072	&	XRT/PC	&	2012-07-10 13:44:54	&	2012-07-10 13:52:46	&	471	\\
00037884073	&	XRT/PC	&	2012-07-12 14:11:08	&	2012-07-12 14:28:53	&	1061	\\
00037884074	&	XRT/PC	&	2012-07-13 13:59:44	&	2012-07-13 14:14:54	&	893	\\
00037884076	&	XRT/PC	&	2012-07-15 17:15:38	&	2012-07-15 17:34:54	&	1143	\\
00037884078	&	XRT/PC	&	2012-07-18 17:30:12	&	2012-07-18 17:44:55	&	870	\\
00037884079	&	XRT/PC	&	2012-07-19 09:23:35	&	2012-07-19 14:20:55	&	993	\\
00037884080	&	XRT/PC	&	2012-07-20 09:32:19	&	2012-07-20 23:55:56	&	364	\\
00037884081	&	XRT/PC	&	2012-07-22 01:30:28	&	2012-07-22 02:11:54	&	880	\\
00037884083  &      XRT/PC   &       2012-07-26 01:49:11     &       2012-07-26 03:32:54     &       1209    \\
00037884084  &      XRT/PC   &       2012-07-28 16:26:24     &       2012-07-28 16:42:54     &       985     \\
00037884085	&	XRT/PC	&	2012-08-01 02:02:36	&	2012-08-01 02:18:54	&	973	\\
00037884087  &     XRT/PC    &      2012-08-07 17:04:36     &       2012-08-07 17:23:55     &       1143    \\
00037884088  &     XRT/PC    &       2012-08-09 15:42:19     &       2012-08-09 15:56:54     &       860           \\
00037884089  &     XRT/PC   &       2012-08-11 07:59:36     &       2012-08-11 14:28:56     &       913         \\
  \noalign{\smallskip} 
  \hline
  \end{tabular}
  \end{center}
  \end{table*} 

\end{document}